\documentclass[twocolumn,showpacs,amsmath,amssymb]{revtex4}
\usepackage{amsmath}
\usepackage{graphicx}
\usepackage{amsfonts}
\usepackage{dcolumn}
\usepackage{bm}

\usepackage{CJK}

\begin{document}
\title{Explosive synchronization transitions in complex neural networks}

\author{Hanshuang Chen$^{1,2}$} 

\author{Gang He$^1$}

\author{Feng Huang$^3$}

\author{Chuansheng Shen$^{4}$}

\author{Zhonghuai Hou$^{2}$}\email{hzhlj@ustc.edu.cn}

\affiliation{$^{1}$School of Physics and Material Science, Anhui
University, Hefei, 230039, People's Republic of China  \\
$^{2}$Hefei National Laboratory for Physical Sciences at Microscales
\& Department of Chemical Physics, University of
 Science and Technology of China, Hefei, 230026, People's Republic of China \\
 $^3$Department of Mathematics and Physics, Anhui University of Architecture, Hefei 230601,
People's Republic of China \\
 $^{4}$Department of Physics, Anqing Teachers College, Anqing, 246011, People's Republic of China}

\date{\today}

\begin{abstract}
It has been recently reported that explosive synchronization
transitions can take place in networks of phase oscillators
[G\'omez-Garde\~nes \emph{et al.} Phys.Rev.Letts. 106, 128701
(2011)] and chaotic oscillators [Leyva \emph{et al.} Phys.Rev.Letts.
108, 168702 (2012)]. Here, we investigate the effect of a
microscopic correlation between the dynamics and the interacting
topology of coupled FitzHugh-Nagumo oscillators on phase
synchronization transition in Barab\'asi-Albert (BA) scale-free
networks and Erd\"os-R\'enyi (ER) random networks. We show that, if
the width of distribution of natural frequencies of the oscillations
is larger than a threshold value, a strong hysteresis loop arises in
the synchronization diagram of BA networks due to the positive
correlation between node degrees and natural frequencies of the
oscillations, indicating the evidence of an explosive transition
towards synchronization of relaxation oscillators system. In
contrast to the results in BA networks, in more homogeneous ER
networks the synchronization transition is always of continuous type
regardless of the width of the frequency distribution. Moreover, we
consider the effect of degree-mixing patterns on the nature of the
synchronization transition, and find that the degree assortativity
is unfavorable for the occurrence of such an explosive transition.

\end{abstract}
 \maketitle

\section{Introduction}
Synchronization is an emerging phenomenon of an ensemble of
interacting dynamical units that is ubiquitous in nature, such as
neurons, fireflies or cardiac pacemakers
\cite{Winfree1990,Strogatz2003,Pikovsky2003}. Inspired by the
seminal works of small-world networks by Watts and Strogatz
\cite{Nature.393.440} and scale-free networks by Barab\'asi and
Albert \cite{Science.286.509}, synchronization on complex networks
has been widely studied \cite{PRP06000175,PRP08000093,RMP08001275}.
These studies have revealed that the topology of a network has
strong influences on the onset of synchronization
\cite{PhysRevE.65.026139,EPL68.603,PhysRevE.72.026208,Chaos,PhysRevLett.96.114102},
path towards synchronization \cite{PhysRevLett.98.034101}, and the
stability of the fully synchronized state
\cite{PhysRevLett.80.2109,PhysRevLett.89.054101,PhysRevLett.91.014101,PhysRevLett.96.034101}.
However, the continuous nature of the synchronization phase
transition is not affected by the topology of the underlying
network, even in heterogeneous scale-free networks.

Recently, explosive transition in complex networks has received
considerable attention since the discovery of an abrupt percolation
transition in random networks
\cite{Science323.1453,PhysRevLett.103.255701} and scale-free
networks \cite{PhysRevLett.103.168701,PhysRevLett.103.135702},
although some later studies claimed this transition is actually
continuous but with unusual finite size behavior
\cite{PhysRevLett.105.255701,PhysRevLett.106.225701,Science333.322}.
Along this line, very recently it was shown that explosive
synchronization transitions can take place in scale-free networks of
phase oscillators \cite{PhysRevLett.106.128701} and chaotic
oscillators \cite{PhysRevLett.108.168702}. The mechanism responsible
for such discontinuous synchronization transitions is the presence
of a positive correlation between the heterogeneity of the
connections and the natural frequencies of the oscillators. These
studies open new perspectives on the research of synchronization
transitions of other dynamical systems. However, the research on
this topic is still in its infancy and deserves more investigations.

In the present paper, we investigate the criticality of phase
synchronization transition of coupled FitzHugh-Nagumo (FHN)
oscillators in Barab\'asi-Albert (BA) scale-free networks
\cite{Science.286.509} and Erd\"os-R\'enyi (ER) random networks
\cite{Publ.Math.6.290,PhysRevE.64.026118}. It is shown that the
positive correlation between degrees and natural frequencies of
nodes can lead to a clear hysteresis loop in synchronization diagram
of BA networks and thus signals the occurrence of an explosive
synchronization transition in heterogeneous networks. And a
sufficient wide distribution of natural frequencies of the
oscillators is necessary to induce such an explosive transition in
BA networks, while in more homogenous ER networks the
synchronization transition is always continuous no matter how wide
the distribution is. Furthermore, we consider the effect of
degree-mixing patterns on the nature of the synchronization
transition. We find that the degree assortativity is unfavorable for
the occurrence of such an explosive transition.

\section{Model}
Let us consider a network of $N$ coupled non-identical FHN
oscillators, a representative model of excitable systems such as
neurons, wherein the dynamics of the $i$th oscillator is described
by the following equations \cite{PhysRevLett.78.775},
\begin{eqnarray}
   \varepsilon   \dot x_i  &=& x_i  - x_i^3  - y_i  + C\sum\limits_{j = 1}^N {A_{ij} \left( {x_j  - x_i }
   \right)},
  \\
   \dot y_i & =& x_i + a_i + \xi _i(t).
\end{eqnarray}
The two dimensionless variables $x$ and $y$ are a voltage-like and a
recovery-like variable, or in the terminology of physical chemistry
and semiconductor physics, an activator and an inhibitor variable,
respectively. The time scale ratio $\varepsilon$ is much smaller
than one (we here set $\varepsilon=0.01$), implying $x$ is the fast
and $y$ is the slow variable. The parameter $a_i$ describes the
excitability of the $i$th unit. If $|a_i|>1$, the system is
excitable, while $|a_i|<1$ implies that the system is oscillatory.
The element of adjacency matrix of the network takes $A_{ij}=1$ if
nodes $i$ and $j$ are connected, and $A_{ij}=0$ otherwise. $C$ is
the coupling constant, and $\xi _i(t)$ is Gaussian noise that is
independent for different units and satisfies $\left\langle {\xi _i
(t)}\right\rangle  = 0$ and $\left\langle {\xi _i (t)\xi _j
(t')}\right\rangle = 2D\delta _{ij} \delta (t - t')$ with noise
intensity $D$.

To establish a microscopic correlation between the dynamics and the
topological properties of nodes, we assume that the natural
frequency $\omega_i$ of the $i$th unit is an increasing function of
its degree $k_i$, where $k_i  = \sum\nolimits_j {A_{ij} }$ is the
number of nodes that are adjacent to the node. Since the oscillation
frequency is a decreasing function of $a_i$, we thus consider $a_i$
decreases linearly with $k_i$ for simplicity, i.e.,
\begin{eqnarray}
a_i  = 0.99 - \delta \frac{{k_i  - k_{\min } }} {{k_{\max }  -
k_{\min } }},
\end{eqnarray}
where $k_{max}$ and $k_{min}$ are the maximum and minimum degree in
the network, respectively. The factor $\delta$ determines the slope
of the linear distribution. The larger $\delta$ is, the wider
distribution of $a_i$ has, or equivalently, a wider frequency
distribution. Other parameters $N=200$, $D=0.005$, and the average
degree $\left\langle k \right\rangle=6$ are fixed in this paper
unless otherwise specified.

To characterize synchronization behavior among the $N$ oscillators,
we first define the phase of the oscillator $i$ as
\cite{PhysRevE.61.R1001}
\begin{eqnarray}
\phi _i (t) = 2\pi \frac{{t - \tau _i^k }} {{\tau _i^k  - \tau _i^{k
+ 1} }} + 2\pi k,
\end{eqnarray}
where $\tau_i^k$ is the time of the $k$th firing of the oscillator
$i$, which is defined in simulation by threshold crossing of
$x_i(t)=1.0$. Thus, the degree of phase synchronization can be
measured by calculating $r = \left\langle {\left| {\frac{1}
{N}\sum\nolimits_{j = 1}^N {e^{i\phi _j } } } \right|}
\right\rangle$, where the vertical bars denote the module and the
angle brackets a temporal averaging. For completely unsynchronized
motion $r\simeq0$, while for fully synchronized state $r\simeq1$.

\begin{figure}
\centerline{\includegraphics*[width=1.0\columnwidth]{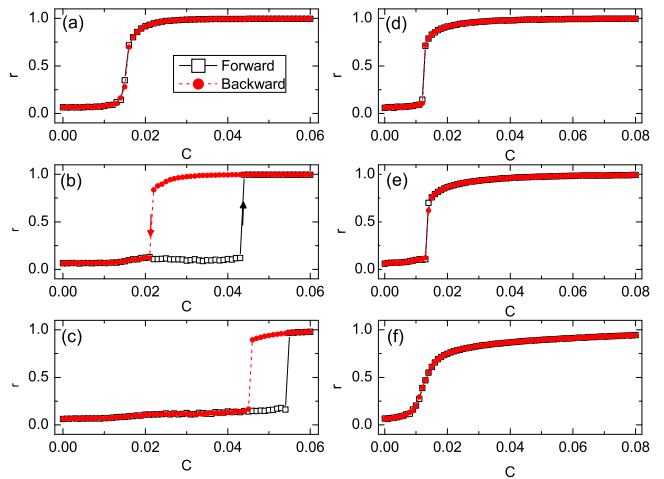}}
\caption{ (Color online) Phase synchronization degree $r$ as a
function of the coupling strength $C$ for different $\delta$ in BA
scale-free networks (left panels) and ER random networks (right
panels). (a) and (d) for $\delta=0.1$, (b) and (e) for $\delta=0.3$,
and (c) and (f) for $\delta=0.9$. Squares (circles) in (a)-(f) mark
the forward (backward) simulations, as $C$ is increased (decreased)
in steps of $C=0.001$. Other parameters are $N=200$,
$\varepsilon=0.01$, $D=0.005$, and the average degree $\left\langle
k \right\rangle=6$. \label{fig1}}
\end{figure}

\section{Results}
The synchronization diagram is obtained by performing both forward
and backward simulations. The former is done by calculating
stationary value of $r$ as varying $C$ from $0$ to $0.06$ in steps
of $0.001$, and using the final configuration of the last simulation
run as the initial condition of the next run, while the latter is
performed by decreasing $C$ from $0.06$ to $0$ with the same step.
Fig.1 shows the results of $r$ as a function of $C$ for different
values of $\delta$ in BA scale-free networks (left panels) and in ER
random networks (right panels). For BA networks, one can find that
the nature of synchronization transition drastically changes with
$\delta$. For $\delta=0.1$ (Fig.\ref{fig1}(a)), the results on the
forward and backward simulations coincide, implying that the
synchronization transition is continuous and of a second-order type.
Increasing $\delta$ to $\delta=0.3$ (Fig.\ref{fig1}.(b)), one can
see that as $C$ increases $r$ abruptly jumps from $r\simeq0$ to
$r\simeq1$ at $C=0.044$, which show that a sharp transition takes
place for the onset of synchronization. On the other hand, the
curves corresponding to the backward simulations also shows a sharp
transition from the synchronized state to the incoherent one at
$C=0.021$. The two sharp transitions occur at different values of
$C$, leading to a strong hysteresis loop with respect to the
dependence of $r$ on $C$. Such a feature indicates that an explosive
first-order synchronization transition arises in BA networks due to
the positive correlation between degrees and natural frequencies.
Further increasing $\delta$ to $\delta=0.9$ (Fig.\ref{fig1}(c)), the
nature of first-order phase transition is still present, but the
area of the hysteresis loop becomes smaller. While for ER random
networks (Fig.\ref{fig1}(d-f)), the forward and backward simulations
always coincide regardless of $\delta$, and thus such a correlation
does not induce an explosive synchronization transition in ER
networks. Therefore, a wide enough distribution of natural
frequencies and degree heterogeneity are both necessary ingredients
for the occurrence of such an explosive synchronization transition.

\begin{figure}
\centerline{\includegraphics*[width=0.8\columnwidth]{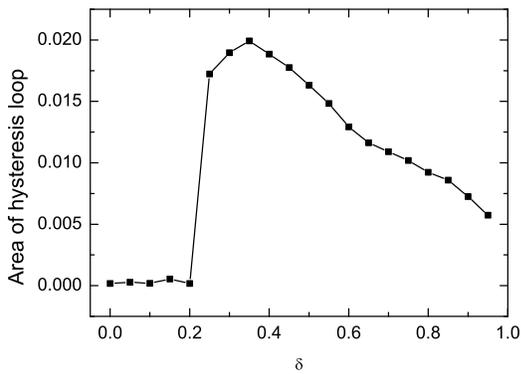}}
\caption{ Area of hysteresis loop in the synchronization diagram of
BA scale-free networks as a function of $\delta$. Other parameters
are the same as Fig.\ref{fig1}. \label{fig2}}
\end{figure}

To show how the criticality of synchronization transition in BA
networks changes with $\delta$, we calculate the area of the
hysteresis loop in synchronization diagram as a function of
$\delta$, as shown in Fig.\ref{fig2}. One can see that this area
equals to zero when $\delta \leq0.2$, implying that the
synchronization transition is of second-order. When $\delta$ is
increased to $\delta=0.25$, the value of this area drastically
changes to a non-zero value. That is to say, the criticality of
synchronization transition changes from a second-order type to a
first-order one at between $\delta=0.2$ and $\delta=0.25$. With
further increasing $\delta$, this area show a nonmonotonic
dependence on $\delta$ and a maximum area occurs at $\delta=0.35$.
This is consistent with the observation in Fig.\ref{fig1}.

\begin{figure}
\centerline{\includegraphics*[width=0.8\columnwidth]{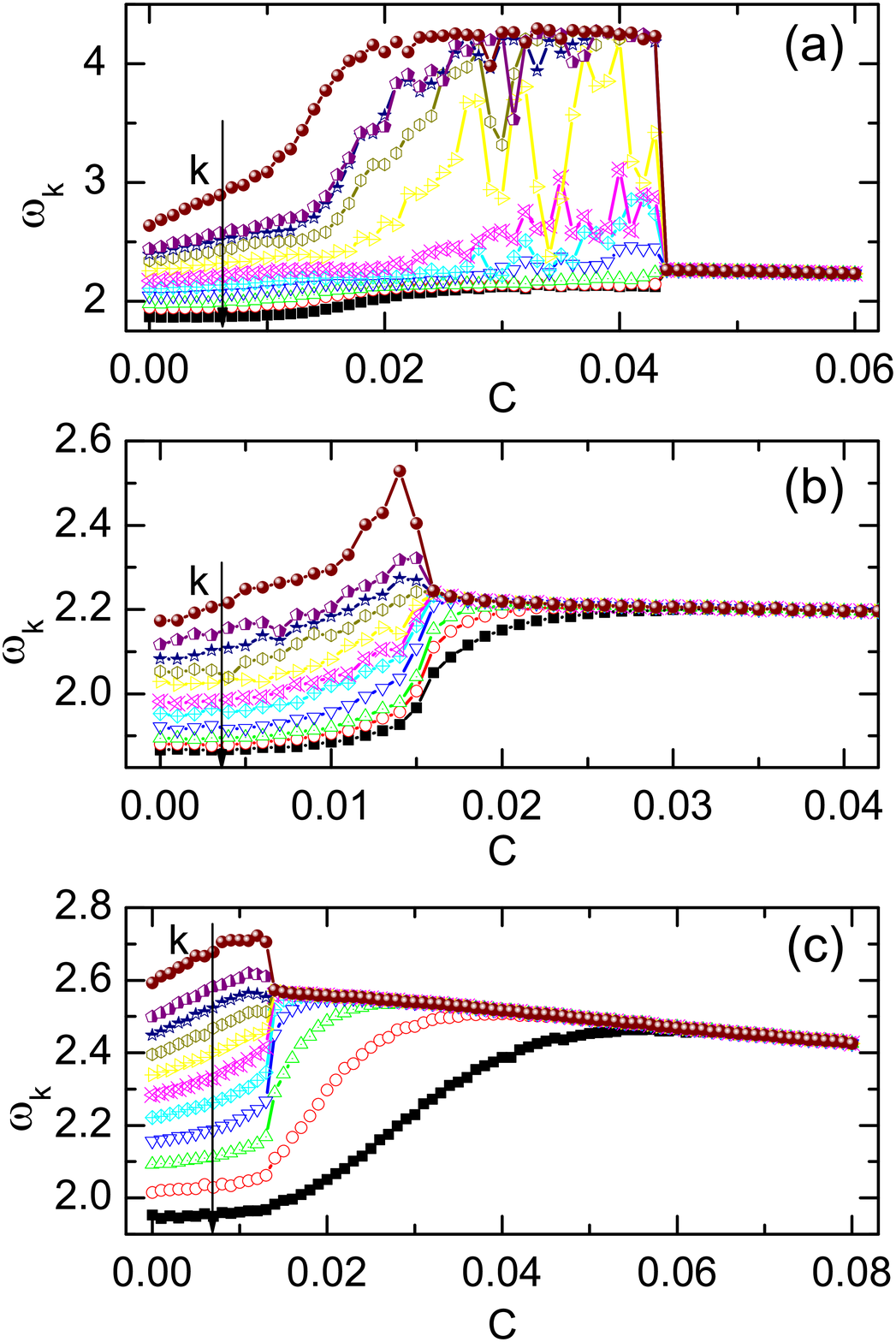}}
\caption{(Color online) The average frequencies $\omega_k$ for
different degrees $k$ along the forward simulation. (a) BA networks:
$\delta=0.3$; (b) BA networks: $\delta=0.1$; (c) ER networks:
$\delta=0.3$. The arrows in (a-c) indicate the decreasing order of
degrees. Other parameters are the same as Fig.\ref{fig1}.
\label{fig3}}
\end{figure}

To further analyze the change of the order of the synchronization
transition, we have calculated the average frequencies $\omega_k$ of
nodes with degree $k$ ($k \in [k_{\min } ,k_{\max } ]$) along the
forward simulation, defined as
\begin{eqnarray}
\omega _k  = \frac{1} {{N_k }}\sum\limits_{i|k_i  = k} {\frac{{\phi
_i (t + T) - \phi _i (t)}} {{2\pi T}}},
\end{eqnarray}
where $N_k$ is the number of nodes that have degree $k$, and $T \gg
1$. By monitoring the variation of $\omega_k$ with $C$ one can
clearly observe that how the full synchronization state is achieved.
In Fig.\ref{fig3}(a) we plot $\omega_k$ as a function of $C$ in BA
networks with $\delta=0.3$. Before the synchronization happens, the
average frequencies $\omega_k$ for nodes with small degrees and
large degrees increase monotonically, while for those nodes with
intermediate degree $\omega_k$ strongly fluctuate between two
transition points. At $C=0.044$, nodes with any degree class
abruptly oscillate synchronously, which signals the explosive
synchronization shown in Fig.\ref{fig1}(b). For comparison, in
Fig.\ref{fig3}(b) and Fig.\ref{fig3}(c) we also plot $\omega_k$ as a
function of $C$ in BA networks with $\delta=0.1$ and in ER networks
with $\delta=0.3$, respectively. One can see that those nodes with
large degree first become locked in frequencies while nodes with the
small degree classes achieve full synchronization later on. This
indicates that the synchronization transition is continuous.

\begin{figure}
\centerline{\includegraphics*[width=0.8\columnwidth]{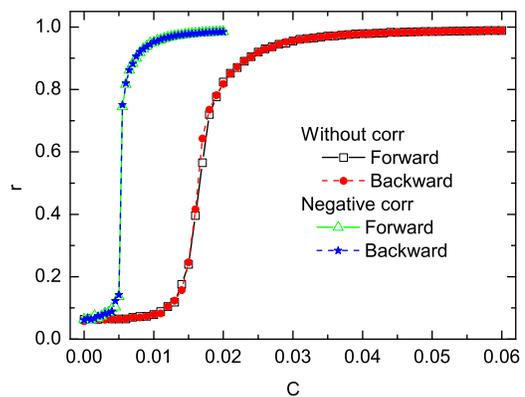}}
\caption{ (Color online) Synchronization diagram in BA scale-free
networks without any correlation and with the negative correlation
between degrees and natural frequencies. Other parameters are the
same as Fig.\ref{fig1}. \label{fig4}}
\end{figure}

Next we will illustrate whether the positive correlation between
degrees and natural frequencies is responsible for the explosive
synchronization. In Fig.\ref{fig4}, we show the results of $r$ as a
function of $C$ in BA networks, where the correlation is destroyed
by randomly shuffle the values of $a_i$. One can see that the
above-mentioned explosive synchronization transition disappears and
the transition becomes a second-order type. Furthermore, we consider
the case of negative correlation between degrees and natural
frequencies. To the end, we set $a_i  = 0.69 + 0.3(k_i - k_{\min
})/(k_{\max}-k_{\min})$ while other parameters keep the same as
Fig.\ref{fig1}(a). The results are also shown in Fig.\ref{fig4} and
the transition is continuous. Therefore, we can safely conclude that
such an explosive synchronization transition arises due to the
positive correlation between degrees and natural frequencies.

\begin{figure}
\centerline{\includegraphics*[width=1.0\columnwidth]{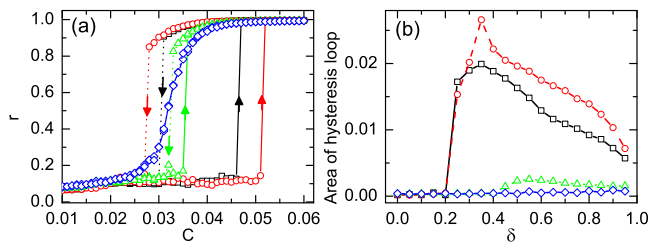}}
\caption{ (Color online) Effect of degree-degree correlations on the
criticality of the synchronization transition. (a) Synchronization
diagram at $\delta=0.5$ for different degree-mixing coefficient
$r_k$. (b) Area of hysteresis loop in synchronization diagram as a
function of $\delta$ for different $r_k$. $r_k=0.0$ (black squares),
$r_k=-0.3$ (red circles), $r_k=0.1$ (green triangles), and $r_k=0.2$
(blue diamonds). Other parameters are the same as Fig.\ref{fig1}.
\label{fig5}}
\end{figure}

Lastly, we will consider the effect of degree-degree correlations on
the criticality of the synchronization transition. It has been
witnessed that many real networks display different degree--mixing
patterns \cite{PRL02208701}. To measure the degree of the
correlation, Newman defined a degree-mixing coefficient as
\cite{PRL02208701}
\begin{eqnarray}
r_k  = \frac{{M^{ - 1} \sum\nolimits_i {j_i k_i  - [M^{ - 1}
\sum\nolimits_i {\frac{1} {2}(j_i  + k_i )} ]^2 } }} {{M^{ - 1}
\sum\nolimits_i {\frac{1} {2}(j_i^2  + k_i^2 )}  - [M^{ - 1}
\sum\nolimits_i {\frac{1} {2}(j_i  + k_i )} ]^2 }},
\end{eqnarray}
where $j_i$ and $k_i$ are the degrees of nodes at the two ends of
the $i$th link with $i = 1, \cdots ,M $ ($M$ is the number of total
links in the network). $r_k=0$ indicates that there is no degree
correlation, while $r_k>0$ ($<0$) indicates that a network is
assortatively (disassortatively) mixed by degree. A previous study
has revealed that degree--mixing pattern plays an important role on
synchronization \cite{PRE06066107}. To generate different
degree--mixing networks, we employ an algorithm proposed in
\cite{PRE04066102}. At each elementary step, two links in a given
network with four different nodes are randomly selected. To get an
assortative network, the links are rewired in such a way that one
link connects the two nodes with the smaller degrees and the other
connects the two nodes with the larger degrees. Multiple connections
are forbidden in this process. Repeat this operation until an
assortative network is generated without changing the node degrees
of the original network. Similarly, a disassortative network can be
produced with the rewiring operation in the mirror method. We start
from BA scale-free networks with a neutrally degree-mixing pattern,
and produce some groups of degree-mixing networks by performing the
above algorithm. Fig.\ref{fig5}(a) plots the synchronization diagram
for different $r_k$ at a fixed $\delta=0.5$. One can see that for
$r_k=-0.3$ the explosive synchronization transition persists. For
$r_k=0.1$ the discontinuous nature of the transition does not
change, but the area of hysteresis loop become rather small. While
for $r_k=0.2$, the forward and backward simulations coincide  and
the synchronization transition becomes continuous. In
Fig.\ref{fig5}(b), we plot the area of hysteresis loop as a function
of $\delta$ for different $r_k$. We find that the case for
$r_k=-0.3$ is similar to that of $r_k=0.0$. For $r_k=0.1$ the area
is larger than zero when $\delta>0.045$ but its value is very small
and has only the order of $10^{-3}$. While for $r_k=0.2$, the area
is always zero regardless of $\delta$, implying the nature of the
synchronization transition has been changed essentially. Therefore,
the degree-mixing patterns have a significant impact on the
criticality of the synchronization transition. For a disassortative
network, the explosive nature of the transition appears if $\delta$
is larger than a threshold value, while for a assortative network
whose degree-mixing coefficient $r_k$ is larger than 0.2, the
explosive transition is absent and the transition becomes
continuous. Since dissortativity implies nodes with larger degrees
tend to connect to those nodes with smaller degrees, local star
configurations are abundant in a disassortative network. Thus, it
seems to suggest that the local star configurations are important
for resulting in such an explosive phenomenon.

\section{Summary}
In summary, we have studied the effect of a microscopic correlation
between degrees and natural frequencies of FHN oscillators on the
criticality of synchronization transition in networks of BA and ER
models. We find that, if the width of frequency distribution is
larger than a critical value, the positive correlation degrees and
oscillation frequencies can lead to the first-order synchronization
transition in BA networks. While in more homogeneous ER network,
such an explosive transition dose ont appear no matter how wide the
distribution is. Therefore, the positive correlation between degrees
and oscillation frequencies and their heterogeneities are both
necessary conditions for such an explosive phenomenon. Moreover, we
have shown the patterns of degree-degree correlations have a
significant impact on the nature of the synchronization transition.
In a disassortative network such an explosive phenomenon persists,
while in an assortative network the transition becomes continuous
type if the degree of assortative correlation is relatively large.
Our results generalize previous findings in phase
\cite{PhysRevLett.106.128701} and chaotic oscillators
\cite{PhysRevLett.108.168702} to relaxation oscillators with
time-scale separation, and suggests that the mechanism for
generating the discontinuous synchronization transition may be
universal. In addition, an explosive transition may imply that the
unsynchronized oscillation state can be metastable with respect to
the full synchronized oscillation state near the synchronization
transition point \cite{Puri2009}. The dynamics of spontaneous
synchronization from the metastable state to the stable full
synchronized state without any change of system parameters may be
interesting and deserves further investigations.

\begin{acknowledgments}
We acknowledge supports from the National Science Foundation of
China (11205002, 91027012, 20933006, and 11147163), ``211 project"
of Anhui University (02303319), and the Key Scientific Research Fund
of Anhui Provincial Education Department (KJ2012A189).
\end{acknowledgments}

%

\end{document}